\documentclass[finalnew]{agujournal2019}
\usepackage{graphicx}

\usepackage{url} 
\usepackage{lineno}
\usepackage{soul}
\usepackage{amsmath}
\usepackage{hyperref} 
\usepackage{natbib}
\usepackage{comment}  
\usepackage{xcolor}
\usepackage{import} 

\nolinenumbers

\draftfalse

\journalname{Geophysical Research Letters}

\begin{document}

\title{
Predictability of Storms in an Idealized Climate Revealed by Machine Learning
}

\authors{Wuqiushi Yao\affil{1,2,3}, Or Hadas\affil{1}\thanks{Co-first author; both co-first authors contribute equally to this paper.}, and Yohai Kaspi\affil{1}}

\affiliation{1}{Dept. of Earth and Planetary Science, Weizmann Institute of Science, Rehovot, Israel}
\affiliation{2}{Key Laboratory of Earth System Numerical Modeling and Application, Institute of Atmospheric Physics, Chinese Academy of Sciences, Beijing, China}
\affiliation{3}{College of Earth and Planetary Sciences, University of Chinese Academy of Sciences, Beijing, China}

\correspondingauthor{Wuqiushi Yao}{yaowuqiushi@gmail.com}
\correspondingauthor{Or Hadas}{or.hadas@weizmann.ac.il}

\begin{keypoints}
\item We use CNN to predict storm tracks skillfully in an idealized GCM, providing a novel framework studying storm predictability.
\item Storm growth is less predictable than displacement, with baroclinicity and jet meanders revealing key source of uncertainties.
\item Explainable AI shows that jet meanders double the sensitivity of the uncertainty in storm intensity to jet structures.
\end{keypoints}

\begin{abstract}
The midlatitude climate and weather are shaped by storms, yet the factors governing their predictability remain insufficiently understood. Here, we use a Convolutional Neural Network (CNN) to predict and quantify uncertainty in the intensity growth and trajectory of over 200,000 storms simulated in a 200-year aquaplanet GCM. This idealized framework provides a controlled climate background for isolating factors that govern predictability. Results show that storm intensity is less predictable than trajectory. Strong baroclinicity accelerates storm intensification and reduces its predictability, consistent with theory. Crucially, enhanced jet meanders further degrade forecast skill, revealing a synoptic source of uncertainty. Using sensitivity maps from explainable AI, we find that the error growth rate is nearly doubled by the more meandering structure. These findings highlight the potential of machine learning for advancing understanding of predictability and its governing mechanisms.

 
\end{abstract}

\section*{Plain Language Summary}

Mid-latitude storms are key drivers of weather and climate variability, yet their predictability remains limited. Using Convolutional Neural Network (CNN) trained on over 100,000 storms from a 200-year idealized climate simulation, we assess the factors controlling forecast accuracy. Our results reveal that storm intensity is significantly harder to predict than storm position, with errors growing fastest in regions of strong vertical wind shear. We further show that a more meandering upper-level jet stream reduces forecast skill by amplifying small initial uncertainties, a factor often overlooked in predictability studies. Using Explainable AI, we pinpoint how storm forecast errors are particularly sensitive to subtle variations in jet structure. These findings highlight the potential of machine learning in diagnosing predictability limits and suggest that integrating AI with traditional dynamical approaches may improve future forecasting skills.

\section{Introduction}

Extratropical cyclones play a crucial role in mid-latitude atmospheric circulation, driving the meridional transport of heat, moisture, and momentum while shaping regional weather patterns and climate variability \citep{ Tamarin2017,Priestley2022}. Understanding their predictability is therefore of critical importance, both in terms of forecasting individual storms and in identifying how the large-scale flow modulates their evolution. From a theoretical perspective, the chaotic nature of the atmosphere implies inherent limits to long-term predictability \citep{Lorenz1963}. This foundational insight, originally derived from idealized models, has driven decades of research into the sources and limits of atmospheric predictability. Previous studies have shown that storms embedded in stronger jet streams tend to intensify more rapidly and exhibit lower predictability \citep{ Froude2007, Pantillon2017, Doiteau2024}. Over Europe, stronger storms have also been linked to greater ensemble spread and reduced forecast skill \citep{Rupp2024}.  These findings align with the theoretical expectations that storms experiencing stronger wind shear (baroclinicity) will grow faster \citep{Eady1949}. Furthermore, \cite{Vallis1983} directly connected increased baroclinicity to reduced predictability using an idealized model.

However, storms evolve within a wide range of climatic and synoptic environments \citep{OrKaspi2025}, which complicates quantifying predictability in the real atmosphere. This complexity obscures the fundamental relationships between background flow patterns and the predictability of individual storms. Idealized general circulation models (GCMs) provide a controlled and statistically stable climate framework that simplifies the dynamics while enabling long integrations and large storm samples \citep{Walker2006,Chemke2015,Tamarin2017,HadasKaspi2021}. The GCMs have therefore played an essential role in predictability studies across a range of climate conditions \protect\citep{Sheshadri2021}.

Even in such simplified frameworks, assessing predictability and its response to perturbations of the initial state often requires large ensemble simulations, which are computationally expensive and complex \citep{DelSole2004,zhang2013sheer,emanuel2016predictability,Coleman2024}. Machine learning offers an efficient alternative by directly learning the mapping between initial conditions and the associated spread of possible outcomes. Recent studies have demonstrated the potential of such probabilistic neural-network frameworks: \citet{Gordon2022} introduced a negative log-likelihood (NLL) loss that allows models to predict both the mean output ($\mu$) and its associated uncertainty ($\sigma$), successfully identifying more predictable initial states, while \citet{brettin2025uncertainty} applied this approach to sea-level forecasts.

Therefore, combining machine learning with idealized GCMs is particularly fruitful: it allows efficient generation of large datasets and the development of accurate, uncertainty-aware models using relatively simple architectures compared with complex operational systems such as Pangu \citep{Bi2023} and GraphCast \citep{Price2025}. The goal of this study is to uncover which aspects of the initial conditions control the forecast uncertainty of midlatitude storms. We train, validate, and test a CNN on 220,000 storm tracks from a 200-year idealized GCM simulation to predict storm displacement and vorticity growth. Section~\ref{sec:data_methods} describes the GCM setup, storm detection, and machine-learning methods. Section~\ref{sec:pred1} assesses the overall predictability of storms, followed by demonstrations of how baroclinicity modulates predictability (Section~\ref{sec:pred2}) and how jet-stream meanders influence forecast uncertainty using an explainable AI technique (Section~\ref{sec:pred3}).

\section{Data and Methods}
\label{sec:data_methods}
\subsection{Idealized GCM}
\label{sec:gcm}
Simulations are conducted using the Idealized Moist Spectral Atmospheric Model model, with a T42 resolution for 200 years \citep{Frierson2006}. The model employs a spectral core that solves the primitive hydrostatic equations for an ideal-gas atmosphere \citep{OGorman2008}. The simulations are carried out in an aquaplanet configuration, where the lower boundary is represented by a slab ocean with a prescribed heat capacity. Further details on the physical process parameterizations are provided in \citep{Tamarin2017}. The climatology of the large-scale circulation is shown in Figure S1, which highlights the subtropical jet (Figure S1a), the eddy-driven jet, and the Hadley cells (Figure S1d). The subtropical jet exhibits stronger vertical shear than the eddy-driven jet near 45$^\circ$N, reflecting differences in baroclinicity between the two jets. Since the aquaplanet setup ensures hemispheric symmetry, data is aggregated from both hemispheres and shown as the northern hemisphere.

\subsection{Tracking Algorithm}
\label{sec:track}
An objective feature-point identification and tracking technique \citep{Hodges1995} is used to detect and track cyclones in the GCM. In this study, we identify cyclones between 20$^\circ $N-60$^\circ $N and between 20$^\circ $S-60$^\circ $S using the 838\,hPa vorticity field, with their centers tracked every 6 hours. A minimum vorticity threshold of $10^{-5} \ \text{s}^{-1}$ is applied, and only cyclones that travel more than 1000 km and persist for over 2 days are included in the analysis. About 220,000 cyclones are identified within a 200 years' run of GCM. The tracks are also used to produce cyclones' initial conditions. For each cyclone identified, we place a box around its genesis location extending $\pm$30$^\circ$ and $\pm$60$^\circ$ in the meridional and zonal directions, respectively. Then, the initial conditions that the cyclones experience are recorded, namely temperature, zonal wind, meridional wind, and specific humidity on 16 vertical levels, extending from 225\,hPa to 963\,hPa. The codes of storm tracking from GCMs and postprocessing for running the CNN are shown in \cite{Yao2025dataset}.

\begin{figure}[htbp]
    \centering
    \includegraphics[width=1.0\textwidth]{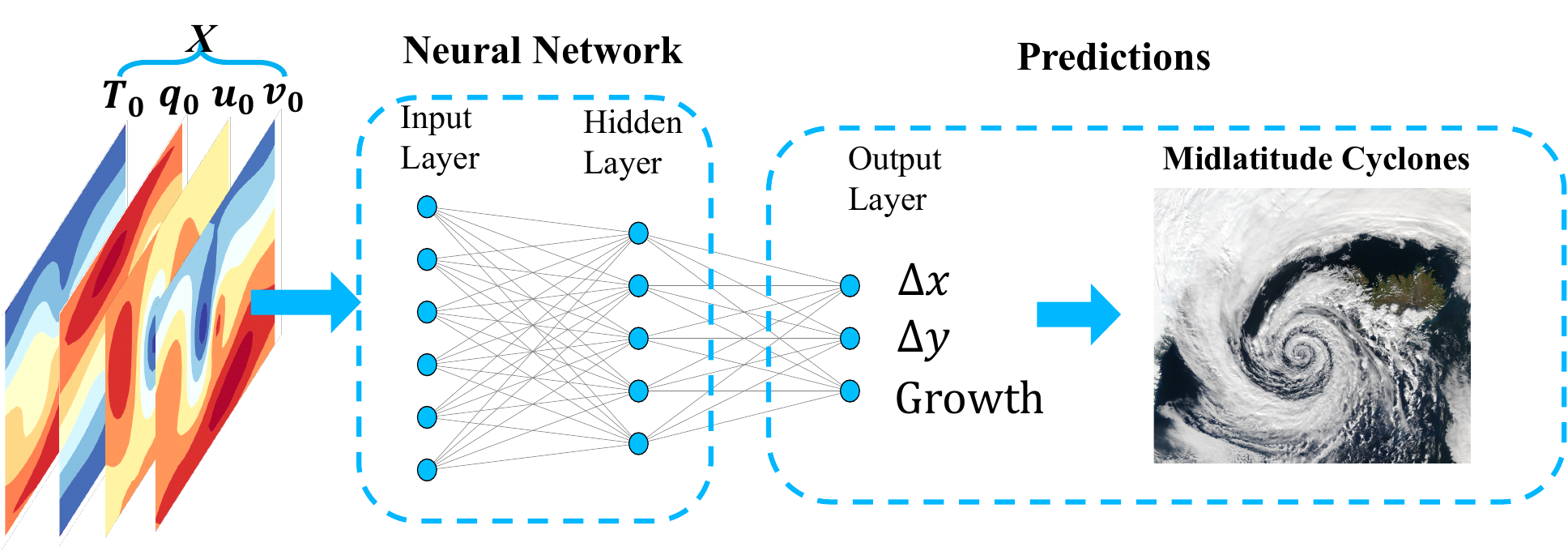}  
    \caption{A schematic diagram illustrating the input and output structure of the neural network. The model takes atmospheric conditions around the storm at genesis as input and predicts storm growth, $\Delta x$ and $\Delta y$ (defined in Section~\ref{sec:machine}) up to 42 hours ahead in 6-hour intervals.}
     
\end{figure}

\subsection{Machine Learning Model}
\label{sec:machine}

In this study, we train a machine learning model to predict the intensity change, zonal and meridional displacements (hereafter, growth, $\Delta x$ and $\Delta y$) of storm-center 838 hPa vorticity anomalies using the atmospheric state around the storm at genesis as input (Figure 1; Section\protect~\ref{sec:track}). All input and out variables are normalized by subtracting the ensemble mean across all storms and dividing by the corresponding standard deviation at each grid point. (see Table S1 for the mean and standard deviation of output variables.)

The primary machine learning architecture employed in this study is a convolutional neural network (CNN). The CNN consists of six convolutional layers with hidden channel sizes of [92, 184, 184, 368, 368, 732], followed by three fully connected layers of sizes [512, 256, 16]. Each convolutional layer uses a kernel size of 3 with padding of 1. Except for the first layer (stride 1), a stride of 2 is applied, halving the spatial dimensions of the input maps at each layer. A dropout rate of 30\% is applied after each convolutional layer to prevent overfitting, and ReLU is used as the activation function throughout. The model is trained using the Adam optimizer with a learning rate of $1 \times 10^{-4}$. For validation, we additionally implement a dense neural network (DNN) and a Linear Regression. Among these models, the CNN consistently achieves the best performance (Figure S2), while Linear Regression shows the worst performance and nearly no skill in predicting the storms. Thus, CNN serves as the basic model of this paper. This study utilizes 77,000 cyclones from the idealized GCM output for model training, 33,000 for validation, and 110,000 for testing. A large test set is employed to ensure the robustness of the analysis. All results shown are based on the "test" dataset. The analysis focuses on the first 42-hour forecasts, during which the model exhibits its highest skill. All codes regarding the model are included in \cite{Yao2025code}. 

In this study, Negative Log-Likelihood (NLL) loss function \citep{Guillaumin2021,Barnes2021} has been adopted, which allows models to jointly predict both the target variable and its associated uncertainty:
\[
\mathcal{L}_{\boldsymbol{\theta}}(\boldsymbol{\chi}) =  \frac{1}{2} \log 2\pi \boldsymbol{\sigma}_{\boldsymbol{\theta}}^2(\boldsymbol{\chi}) + \frac{[\mathbf{f}(\boldsymbol{\chi}) - \boldsymbol{\mu}_{\boldsymbol{\theta}}(\boldsymbol{\chi})]^2}{2 \boldsymbol{\sigma}^2_{\boldsymbol{\theta}}(\boldsymbol{\chi})}, 
\tag{1}\]
where $\mathcal{L}$ is the NLL loss; {$\mathbf{\chi}$} is the model input; {$\mathbf{f}$} is the ground truth of the output— growth, $\Delta x$, and $\Delta y$ {estimated from the 6th to 42nd hours at a 6-hour interval in this study, therefore $\mathbf{f}$ is a vector of (7,3)}; and $\boldsymbol{\theta}$ represents model's learnable parameters.  In this framework, each output is treated as a Gaussian probability distribution, where $\boldsymbol{\mu}$ and $\boldsymbol{\sigma}$, respectively, represent the center and spread of the Gaussian. The NLL loss acts to maximize the Gaussian likelihood. Specifically, the formulation allows the model to reduce loss in multiple ways: by predicting a mean ($\boldsymbol{\mu}$) close to the target, by assigning a higher variance ($\boldsymbol{\sigma}^2$) to less predictable inputs, or both \citep{Gordon2022}. As a result, high-error predictions are not penalized as long as they are accompanied by appropriately high predicted uncertainty, encouraging the model to learn and express initial-state-dependent predictability through the variable $\boldsymbol{\sigma}^2$. {In practice, for the optimization, the $\mathcal{L}$ is averaged over all outputs and all time steps:}

\[\mathcal{J}(\boldsymbol{\theta})
= \frac{1}{N K} \sum_{i=1}^{N} \sum_{k=1}^{K}
\frac{1}{2} \left[
\log\!\big(2\pi \sigma_{\boldsymbol{\theta},k}^2(\boldsymbol{\chi}_i)\big)
+ \frac{
\big(f_k(\boldsymbol{\chi}_i)
- \mu_{\boldsymbol{\theta},k}(\boldsymbol{\chi}_i)\big)^2
}{
\sigma_{\boldsymbol{\theta},k}^2(\boldsymbol{\chi}_i)
}
\right],\tag{2}\]
 {where N denotes the number of samples, and K is the number of components(7$\times$3) in the predicted $\mu$ and $\sigma^2$.}  {Notably, our results show that the predicted variance $\boldsymbol{\sigma}^2$ is consistent with the realized MSE, indicating that the network provides a statistically calibrated estimate of forecast uncertainty (Figure~S3; Text S1).}

\subsection{Explainable AI}
\label{sec:xai}
In order to quantify the weight of each input variable in the forecast uncertainty, we apply sensitivity analysis \citep{Simonyan2013}, a gradient-based method widely used in machine learning to attribute model outputs to inputs. It allows to identify which variables dominate the prediction and to assess whether forecast errors or uncertainties are linked to these influential inputs. We define the sensitivity of the predicted mean of Gaussian as:
\[
\mathcal{S}(\boldsymbol{\chi}) = \left| \frac{\partial \boldsymbol{\mu}_{\boldsymbol{\theta}}(\boldsymbol{\chi})}{\partial \boldsymbol{\chi}} \right|,
\tag{3}
\]
where {$\mathcal{S}(\mathbf{\chi})$} measures the impact of small input perturbations on the predicted mean $\boldsymbol{\mu}$ (Equation 1). Since the NLL loss optimizes both the center and the spread of a Gaussian, we can also calculate the sensitivity of this uncertainty spread to the input:

\[
\mathcal{E}(\boldsymbol{\chi}) = \left| \frac{\partial \boldsymbol{\sigma}_{\boldsymbol{\theta}}^2({\boldsymbol{\chi}})}{\partial \boldsymbol{\chi}} \right|,
\tag{4}
\]
where, $\mathcal{E}(\mathbf{\chi})$ quantifies how sensitive uncertainty of the output variables is to some small errors in the initial field $\mathbf{\chi}$ . Together, $\mathcal{S}(\mathbf{\chi})$ and $\mathcal{E}(\mathbf{\chi})$ offer insight into how initial conditions contribute to the output and its uncertainty.  {Since in machine learning models all variables are normalized, the $\mathcal{S}$ and $\mathcal{E}$ have no dimension, which allows comparison.}


\section{Results}

\subsection{The predictability of cyclone tracks}
\label{sec:pred1}


We first examine storm predictability as revealed by the CNN’s forecasts applied to the aquaplanet GCM. For all variables: growth, $\Delta x$ and $\Delta y$, {The forecast error increases monotonically over time (Figure 2a; see Figure S4a for MSE shown in dimensional form). In contrast, the $R^2$ peaks at the 18th hour rather than during the first few hours, likely because storm displacement and growth are too small to capture within the initial 12 hours (Table S1), making these early changes less relevant to the initial state (Figure S4b). However, storm growth remains the most difficult quantity, motivating our focus on growth in the following discussion.}

To explore the predictability of growth, Figure 2b constructs a two-dimensional PDF of the predicted $\boldsymbol{\mu}$ and $\boldsymbol{\sigma}^2$ (abscissa and ordinate, respectively) of the predicted Gaussian distribution (Equation 1) for 42-hour growth. Variance reflects forecast uncertainty, with larger values indicating higher uncertainty. The 70\% maximum density contour forms a tilted ellipse, while the contour connecting the peak density in each bin forms a distinct “V” shape, centered around the point of maximum predicted growth. These patterns indicate that rare events are inherently more difficult to predict  {as expected}. {In addition to this trend,} considerable variability in uncertainty remains even for a given growth rate. To test what drive this variability in predictability, two regions are selected that exhibit similar predicted growth ($0$–$2.0 \times 10^{-5}\ \mathrm{s}^{-1}/42\ \mathrm{h}$) but contrasting levels of uncertainty—one with low (good predictability) and the other with high (poor predictability) $\boldsymbol{\sigma}^2$. Their implications for predictability will be examined in Section~\ref{sec:pred3}.

To uncover which input variables and physical processes contribute most to predictability, Figure 2c presents the sensitivity (Equation 3) of the 42-hour growth prediction to all input variables. Sensitivity peaks at 838\,hPa for all variables, as expected, since this is the storm-tracking level where environmental fields most directly influence vorticity dynamics. Among all variables, winds, which are directly linked to vorticity, contribute most to growth. Zonal wind slightly exceeds meridional wind due to its stronger association with the jet stream that guides storm motion. Mid-tropospheric temperature (500–750\,hPa) surpasses meridional wind, reaching its maximum influence on growth within this layer. {The results above demonstrate that the combination of machine learning and idealized GCM is able to make skillful and uncertainty-aware predictions, and identify which regions and variables contribute most to the predictability. Next, we would investigate what exact patterns are associated with high and lower predictability.}

\begin{figure}[ht]
    \centering
    \includegraphics[width=1.0\textwidth]{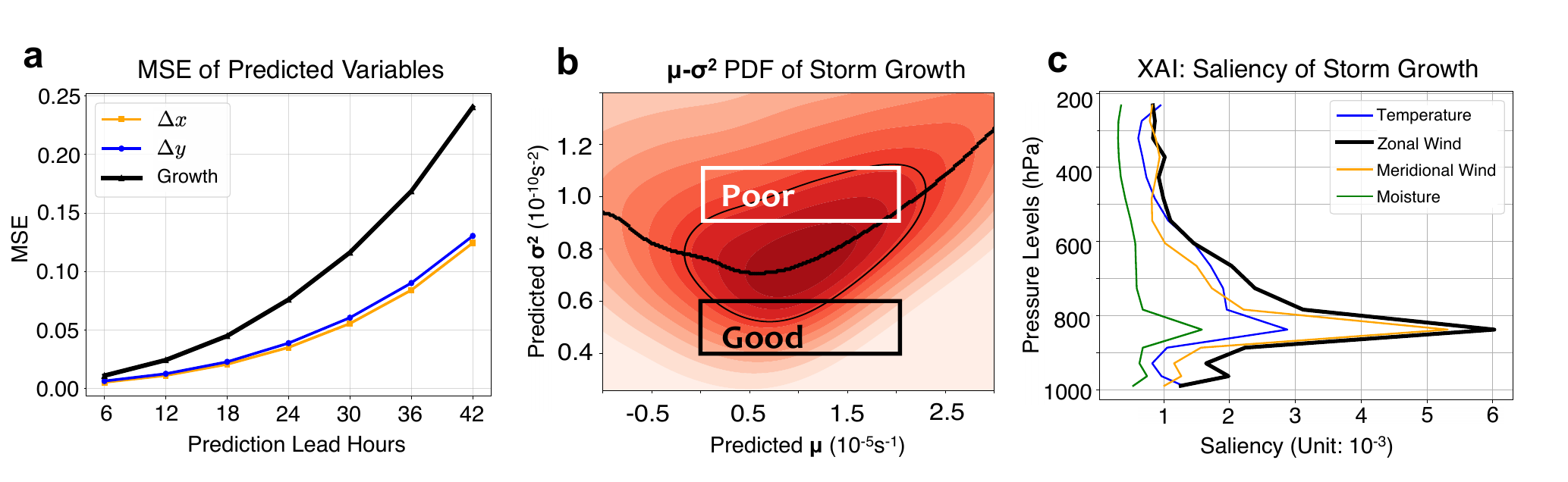}
    \caption{(a) Mean squared error (MSE) normalized by $42^{\mathrm{nd}}$-hour {variance} of the CNN-predicted $\Delta x$, $\Delta y$ and growth .  (b) Joint probability density function (PDF; shading) of the predicted $\boldsymbol{\mu}$ and $\boldsymbol{\sigma}^2$ (defined in Equation 1) from the machine learning models for the $42^{\mathrm{nd}}$-hour growth. The thin black contour encloses 70\% of the maximum density, and the thick contour indicates the peak density within each bin  {of predicted $\boldsymbol{\mu}$}. The two boxes denote regions used to separate good and poor predictions.  {Contours are plotted at intervals of 0.03 from 0.0 to 0.3.}  (c)  {Vertical structure of the sensitivity of the prediction to perturbations in each input variable $\mathcal{S}(\mathbf{x})$ calculated for growth at 42$^{nd}$ hour, averaged horizontally for each variable and pressure level.} 
    }
     
\end{figure}


\subsection{Baroclinicity and predictability}
\label{sec:pred2}

In the aquaplanet simulation, the climate varies primarily with latitude due to the zonally symmetric boundary conditions. This feature allows us to directly relate storm predictability to latitude-dependent background conditions. Figure 3a shows that well-predicted storms preferentially reach peak density near 44$^\circ$N, whereas poorly predicted storms tend to cluster near 34$^\circ$N. To explore the physical origin of this latitudinal dependence, we next examine the background climatology. In this idealized model, the most distinct and dynamically relevant feature of the climatology is the baroclinicity, which varies systematically with latitude. We quantify baroclinicity using the bulk vertical shear, defined as the difference in zonal wind between 275 hPa (model top layer) and 963 hPa (bottom layer). Using Eady growth rate yield qualitatively similar results. The zonal-mean structure (Figure~3b) reveals maximum baroclinicity around $25^\circ$N, associated with the upper-level jet, and a  weakening toward higher latitudes. Notably, this baroclinicity peak lies just south of the latitude band where poorly predicted storms concentrate (around $34^\circ$N), suggesting a possible dynamical link between baroclinicity and forecast uncertainty.

Consistent with \citet{HadasKaspi2024}, storms embedded in more baroclinic regions exhibit faster growth (Figure 3c) and lower predictability, with the MSE of growth increasing with baroclinicity (Figure 3d). Interestingly, the error in $\Delta y$ (propagating 2.8$^\circ$ northward, table S1) decreases with increasing baroclinicity, opposite to storm growth. Our interpretation is that stronger baroclinicity corresponds to enhanced upper-level potential-vorticity gradients, which improve the steering of storms by the large-scale flow \citep{Tamarin2017}, reducing sensitivity to small-scale, less predictable fluctuations. For $\Delta x$ (propagating 20$^\circ$ eastward), this guiding effect seems to be offset by the chaotic background flow, leaving predictability nearly unchanged.

Overall, the systematic alignment between latitude, vertical shear, and forecast skill demonstrates that storms forming in regions of stronger baroclinicity are less predictable in growth, but more predictable in $\Delta y$. Our results provide a clear link between baroclinicity and storm predictability in a medium-complexity aquaplanet model, consistent with findings from more realistic frameworks \citep{Pirret2017}.

\begin{figure}[ht]
    \centering
    \includegraphics[width=1.0\textwidth]{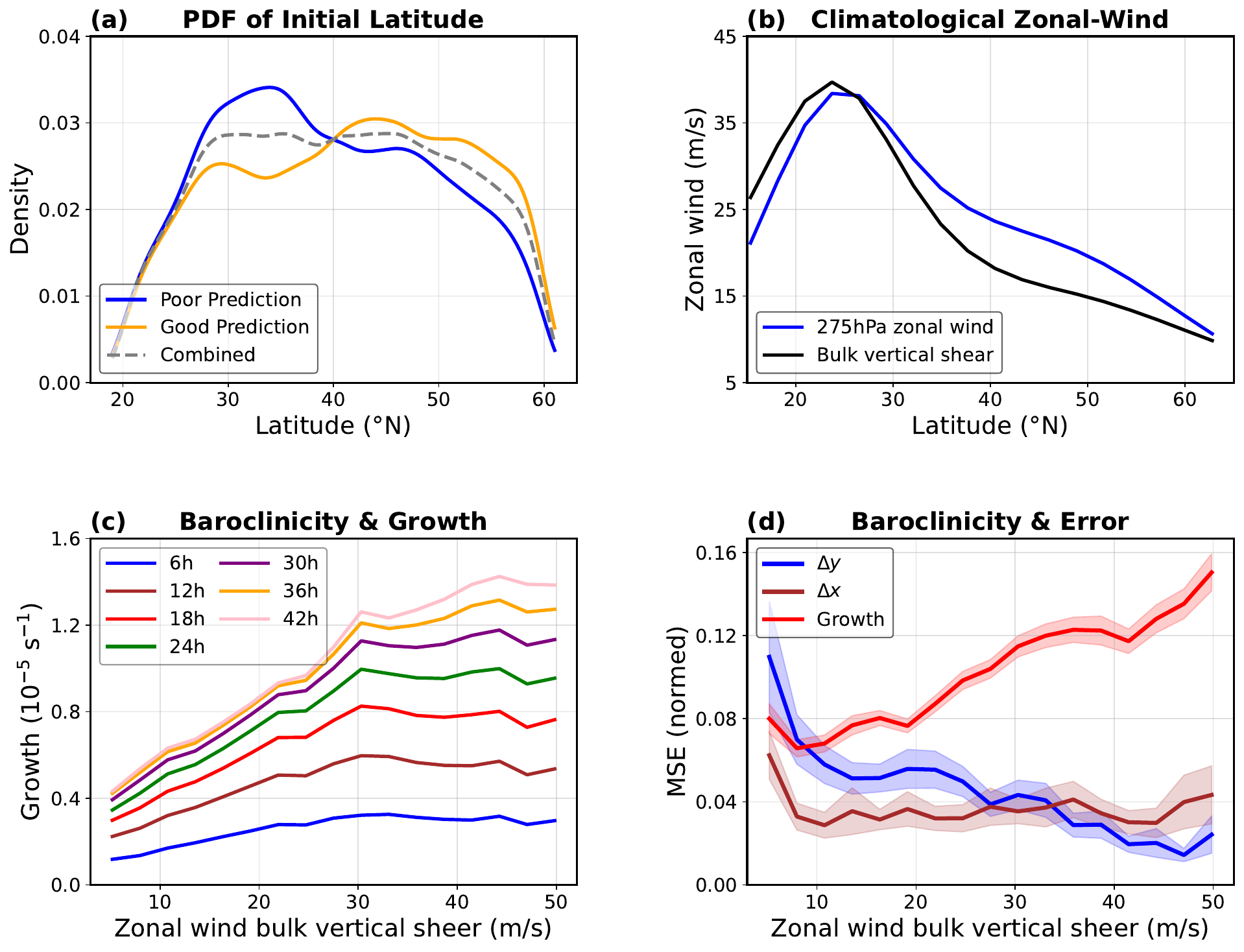}  
    
    \caption{  (a) PDF of the initial storm latitude for well predicted and poorly predicted storms (good and poor prediction, as defined in Figure 2b); (b) Climatological distribution of Zonal wind at 275\,hPa (black) and bulk vertical shear (blue) in the idealized GCM; (c) The mean growth as a function of bulk vertical shear at the storm center, defined in Section~\ref{sec:gcm}, at different lead times; (d) The  {mean} MSE of  $\Delta x$, $\Delta y$ and growth  {averaged over all time steps} as a function of bulk vertical shear. {Shading indicates ±1 standard error, computed as the standard deviation within each bin divided by $\sqrt{N}$. MSE is calculated using variables normalized by the mean and standard deviation (Table S1).};}

\end{figure}

\subsection{Jet meandering, predictability and error growth rate}
\label{sec:pred3}
We next seek to isolate the synoptic-scale conditions that govern forecast predictability. As discussed in Section~\ref{sec:pred2}, climatic conditions vary only with latitude in aquaplanet models. Therefore, separating storms according to latitude removes the climatic variability and leaves only the synoptic variability. To identify the dominant synoptic signal, Figure 4 presents composites of 275\,hPa zonal wind for storms with {"good"} (Figures~4a,b) and {"poor"} (Figures~4d,e) predictability, separately for storms occurring north (Figures~4a,d) and south (Figures~4b,e) of 40$^\circ$N. These structures have been confirmed to be insensitive to sample sizes {(Figure S5)}.

In general, the 275\,hPa zonal wind patterns differ markedly between the good- and poor-prediction groups, both south and north of 40$^\circ$N (Figures 4a,b,d,e). For storms forming south of 40$^\circ$N, the good-prediction group (Figure 4a) exhibits a strong, coherent jet with a 36\,m\,s$^{-1}$ maximum at the storm center and relatively smooth meridional gradients ($\partial u / \partial y > 0$ south of the storm). In contrast, the poor-prediction group (Figure 4d) features a more intense core ($>$40\,m\,s$^{-1}$) but displays strong zonal variability ($\partial u / \partial x$) east of the storm, with a pronounced meandering structure absent in the good-prediction counterpart. North of 40$^\circ$N, both good- and poor-prediction storms are embedded in a jet regime characterized by decreasing wind speeds with latitude ($\partial u / \partial y < 0$), consistent with the climatological background (Figure 3c). The good-prediction group (Figure 4b) shows a relatively smooth jet with only weak meandering north of the storm. In contrast, the poor-prediction group (Figure 4e) reveals strong zonal asymmetry with sharp $\partial u / \partial x$ variations near and east of the storm center, while meridional gradients remain weak. A direct estimate of the relationship between $\partial u / \partial x$ and predictability {(Figure S6)} also supports that such a meandering structure increases forecast uncertainty of growth. Overall, the key distinction lies in the jet structure east of the storm center: poor-prediction storms are consistently associated with more meandering and zonally asymmetric upper-level winds, whereas well-predicted storms are embedded in smoother, more coherent jets. These structures likely modulate downstream uncertainty growth. Similar patterns are observed at lower levels {(Figure S7)}. 

In order to quantify how these structures are linked to predictability, here we evaluate the sensitivity of uncertainty to all the input variables (Equation 4). The results show that forecast uncertainty in 42-hour growth is most sensitive to variations in the zonal-wind structure (Figure S8), highlighting the dominant role of jet dynamics in storm predictability. The “good prediction” group exhibits roughly half the {sensitivity} of the “poor prediction” group to all input variables, indicating a strong correspondence between reduced {sensitivity of uncertainty (to input variables)} and improved forecast skill. To probe the role of jet structure, we compare the sensitivity patterns of storms with well- and poorly- predicted meandering zonal winds (Figures 4c, 4f). Because storms occurring south and north of 40$^\circ$N display similar sensitivity characteristics, we present the composite of them. Both prediction groups show peak sensitivity near the storm center, but the poor-prediction group displays twice the sensitivity. This enhancement suggests that meandering structures east of the storm center substantially accelerate error growth. By contrast, a similar meander 10$^\circ$ north of the center in the good-prediction group (Figures 4a, 4b) contributes little to sensitivity, illustrating that such structures northward of the storm are irrelevant to the storm predictability (Figure 4c). Overall, these findings demonstrate that the spatial configuration of the jet, particularly downstream of the storm center, plays a critical role in shaping the predictability of storm intensity.

\begin{figure}[ht]
    \centering
    \includegraphics[width=1.0\textwidth]{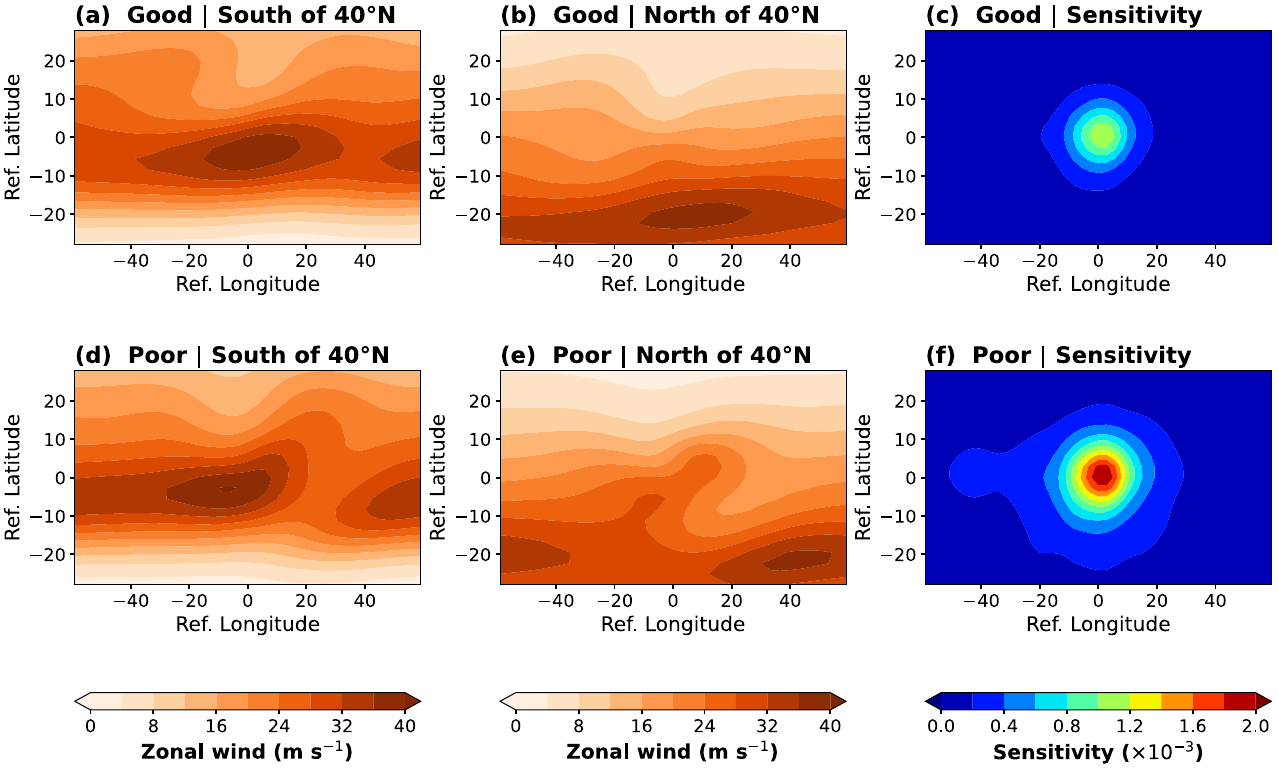}  
    \caption{Composites of 275\,hPa zonal wind fields at storm initialization for (a) well and (d) poorly predicted storms located south of 40$^\circ$ N; (b) and (e) are as (a) and (d), but for storms north of 40$^\circ$ N. The number of samples used for the composites are: (a) 6824; (b) 8677; (d) 7406; and (e) 6353. (c)  {Horizontal distrubution of} $\mathcal{E}(\mathbf{\chi})$ {(equation 4), the derivative of uncertainty in the 42-hour growth to the initial zonal-wind field (unit: 10$^{-3}$), shown} for the good prediction set,  {the sensitivity vector is quantified as the vertical-column mean of the gradient components.}; (f) Same as (c), but for the poor prediction set.}
     
\end{figure}

\section{Conclusion and Discussion}
\label{sec:conclusion}
This study investigates the predictability of mid-latitude storms by applying a convolutional neural network (CNN) to forecast storm tracks in an aquaplanet GCM simulation. By integrating explainable AI with an idealized modeling framework, we find that:

\begin{enumerate}
    \item Storm growth is less predictable than displacement, with large variability across storm samples. Zonal and meridional wind structures contribute most strongly to the predictions. 

    \item  {Stronger baroclinicity enhances storm growth while reducing its predictability, yet simultaneously improves the predictability of meridional displacement.} 

    \item A more meandering jet is linked to decreased predictability of storm growth. Explanable AI results suggest that the more-meandering jet doubles the{sensitivity of uncertainty in storm growth} compared to a less-meandering jet.
\end{enumerate}

These findings demonstrate that combining explainable AI with traditional dynamical analysis can yield deeper insights into the mechanisms governing storm predictability. However, it remains uncertain whether the relationships identified in this idealized setting hold in the real atmosphere. Future work using reanalysis data and operational forecast models is essential to test the generalizability of these findings and evaluate their relevance for real-world weather prediction.

\section*{Conflicts of Interest}

The authors declare no conflicts of interest.

\section*{Open Research Section}

All datasets and code created or analyzed for this study are publicly archived in trusted repositories with persistent identifiers, enabling full transparency, reproducibility, and reuse.

\textbf{Data Availability}  

The data used in this study were generated using the publicly available Geophysical Fluid Dynamics Laboratory (GFDL) Idealized Moist Spectral Atmospheric Model \citep{Frierson2006}, which is fully documented and available at \url{https://www.gfdl.noaa.gov/idealized-moist-spectral-atmospheric-model-quickstart/}. Simulations were performed at T42 resolution for 200 years. 
Due to the large data volume, one representative simulation year, along with configuration files and post-processing scripts, has been archived in Zenodo \citep{Yao2025dataset}.

\textbf{Software Availability}  

The full analysis and machine-learning code, including the CNN architecture, training algorithm, sensitivity-analysis tools, and figure-generation scripts, is archived at Zenodo \citep{Yao2025code}.

\section*{Global Research Statement}

This research did not involve any external collaborators, local institutions, fieldwork, sampling, or community-based data collection. All data were generated via a 200-year aquaplanet experiment using the GFDL model, run on institutional high-performance computing systems. All individuals who met the AGU Publications authorship criteria were included as co-authors; those providing technical support are acknowledged in the Acknowledgements section. No formal permits, agreements with local authorities, or authorizations were required.

In line with AGU’s mandate on “Inclusion in Global Research” and mindful of the TRUST Code’s values of Fairness, Respect, Care, and Honesty, we ensured transparent attribution of contributions and open sharing of data and code. Co-authors jointly participated in the design, implementation, analysis, and interpretation stages of the experiment. The final manuscript was circulated to all co-authors, who verified the results and helped contextualize findings. Data and model scripts have been made publicly available in a trusted repository in accordance with the AGU Data \& Software policy.

Overall, while no cross-regional or community-based collaboration was applicable, we confirm that ethical and scientific considerations have been met in full accordance with AGU policy and the TRUST Code values.

\acknowledgments
W.Y. acknowledges Huayu Chen for inspiring discussions regarding explainable AI. O.H. acknowledges Laure Zanna for meaningful discussions. This research has been supported by the Azrieli fellowship and the Israeli Science
Foundation (Grant 996/20).

%
%


\bibliography{agusample}
\clearpage
\newpage
\section*{Supporting Information}

\setcounter{figure}{0}               
\renewcommand{\thefigure}{S\arabic{figure}}  
\renewcommand{\thetable}{S\arabic{table}}
\noindent\textbf{Contents}
\begin{enumerate}
\item Text S1
\item Table S1
\item Figures S1 to S8
\end{enumerate}

\section*{Text S1. Interpretation of Figure S3}

Figure S3 provides complementary diagnostics for evaluating the probabilistic calibration of the uncertainty-quantifying neural network.  
Panel (a) shows the mean normalized MSE as a function of the predicted variance $\boldsymbol{\sigma}^2$ for each variable ($\Delta x$, $\Delta y$, and growth$)$, with shaded areas representing the $\pm1$ standard error. The overall monotonic relationship confirms that the model’s predicted variances increase proportionally with the realized errors, a necessary condition for statistical consistency. 
Panel (b) relates both quantities to the network confidence percentile $\alpha$. For the relatively predictable variables $\Delta x$ and $\Delta y$, the predicted variances closely follow the realized MSE, indicating sharp and well-calibrated uncertainty estimates. In contrast, for the less predictable growth component, $\hat{\boldsymbol{\sigma}}^2$ systematically exceeds the realized MSE, suggesting that the network adopts a more conservative uncertainty envelope—effectively “covering but not tight.”  \\
Collectively, these diagnostics confirm that the predicted variance $\boldsymbol{\sigma}^2$ faithfully characterizes the underlying forecast uncertainty, serving as a physically meaningful indicator of the system’s intrinsic predictability.

\begin{table}[h]
    \centering
    \includegraphics[width=0.9\textwidth]{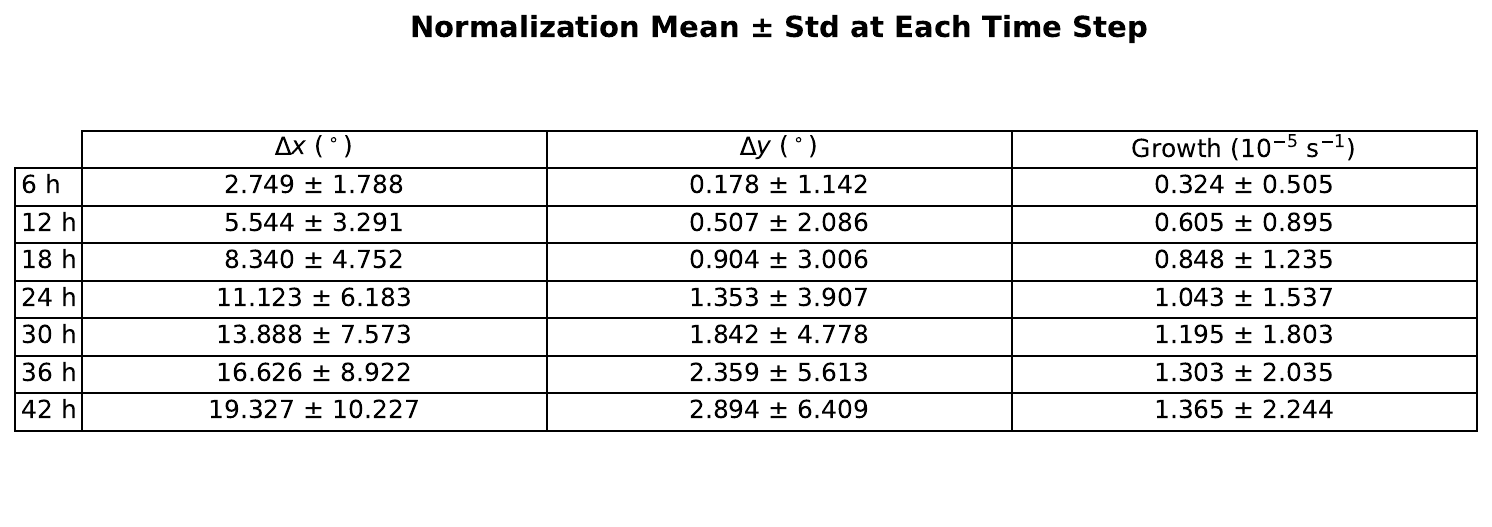}  
    \caption{Normalization mean and standard deviation of the variables: $\Delta x$, $\Delta y$ and  growth used at each time step (Table S1).}
    \label{tab:s1}
\end{table}

\begin{figure}
    \centering
    \includegraphics[width=1.0\textwidth]{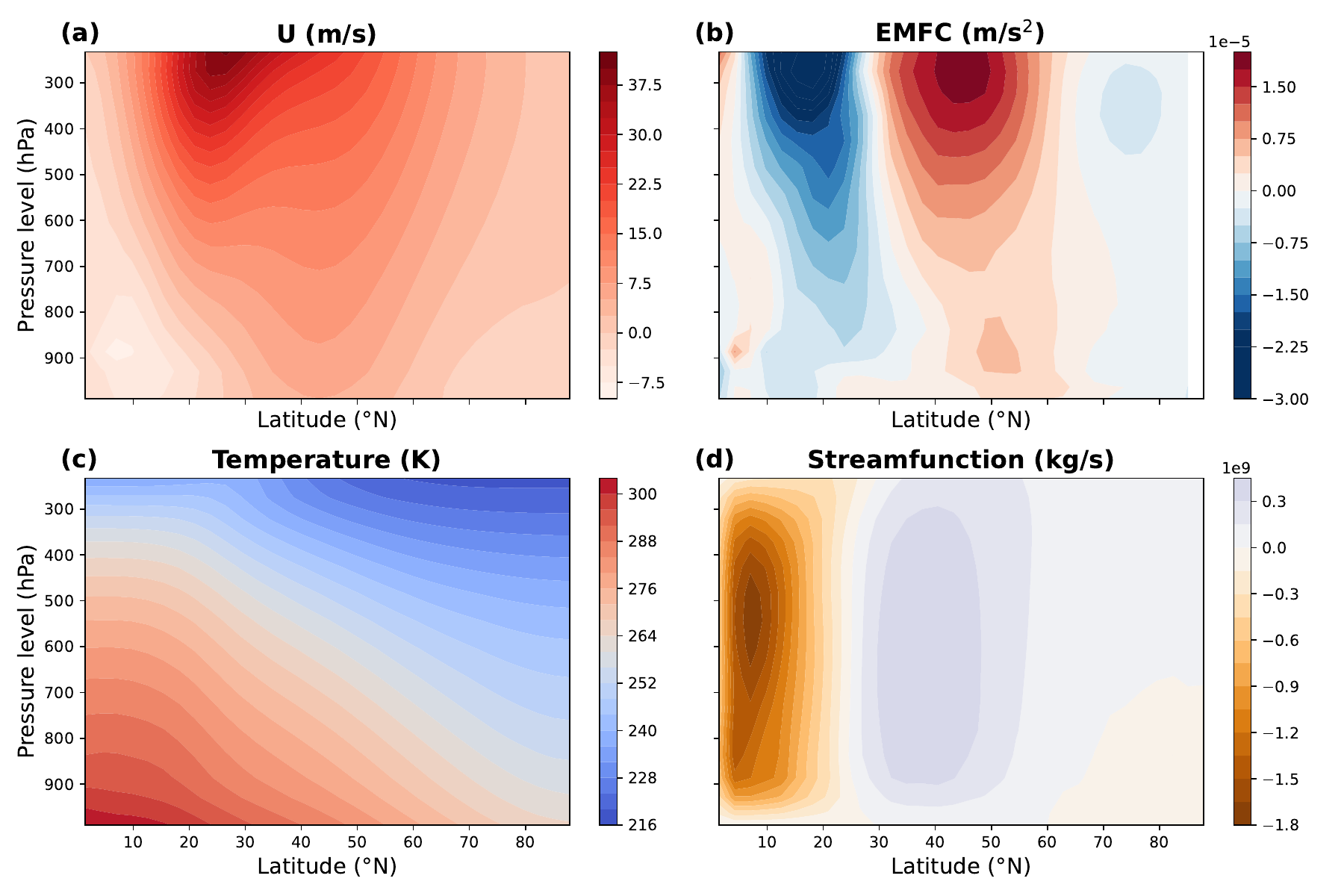}  
    \caption{Zonally averaged climatology of the idealized GCM: (a) Zonal wind(U, $\mathrm{m\,s^{-1}}$); (b) Eddy Momentum Flux Convergence ($-\frac{\partial{\overline{u'v'}}}{\partial y}$, $\mathrm{m\cdot s^{-2}}$, EMFC); (c) temperature (K); (d) Meridional overturning circulation stream function ($\mathrm{kg\cdot s^{-1}}$). }
    \label{fig:s1}
\end{figure} 

\begin{figure}
    \centering
    \includegraphics[width=0.6\textwidth]{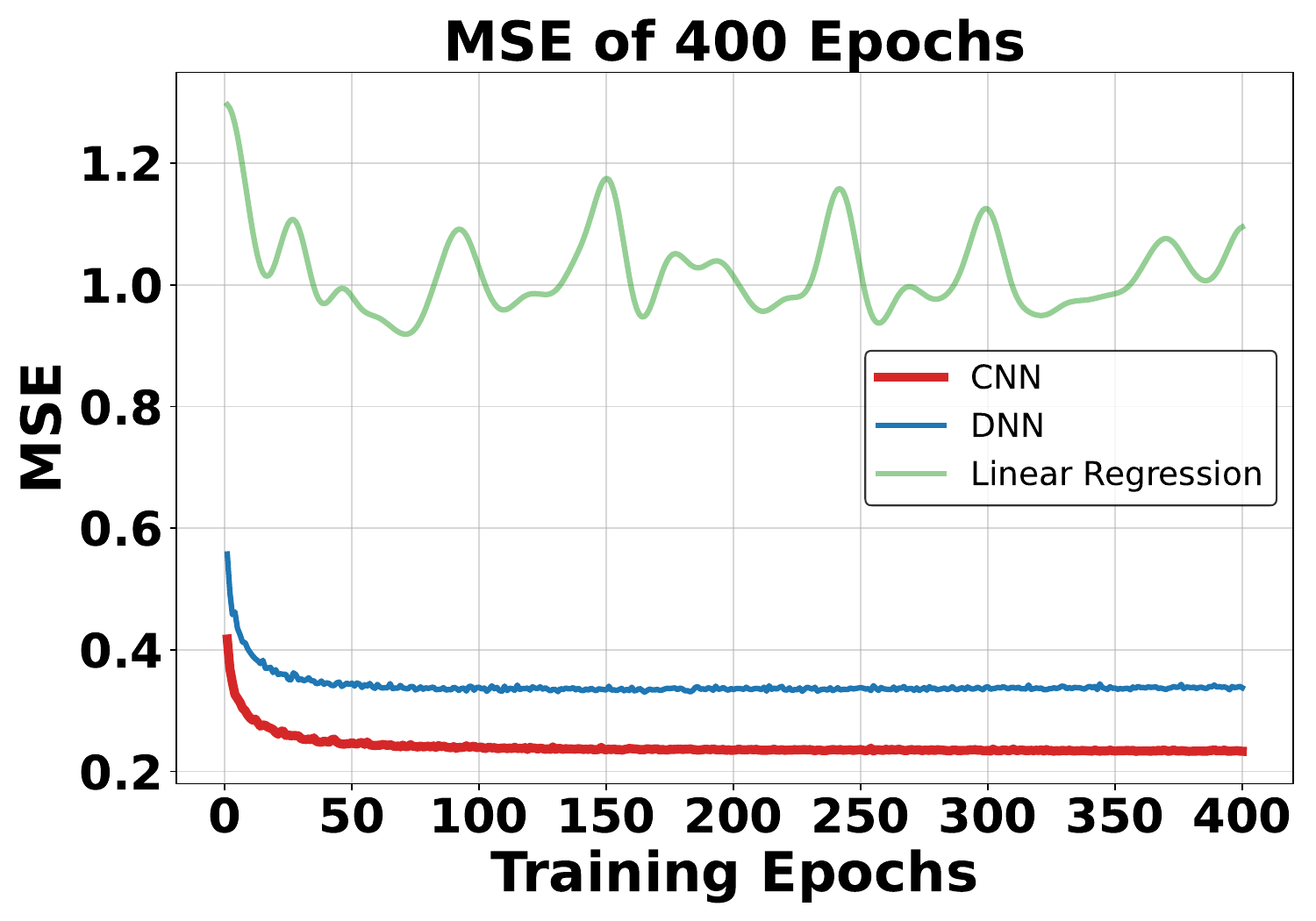}  
    \caption{Comparison of MSE among the three neural-network models predicting storms in this study. The figure shows how the prediction error evolves over 400 training epochs, illustrating the skill achieved by each model. MSE is computed and averaged across all variables, each normalized by its mean and standard deviation (Table S1).}
    \label{fig:s2}
\end{figure}

\begin{figure}
    \centering
    \includegraphics[width=1.0\textwidth]{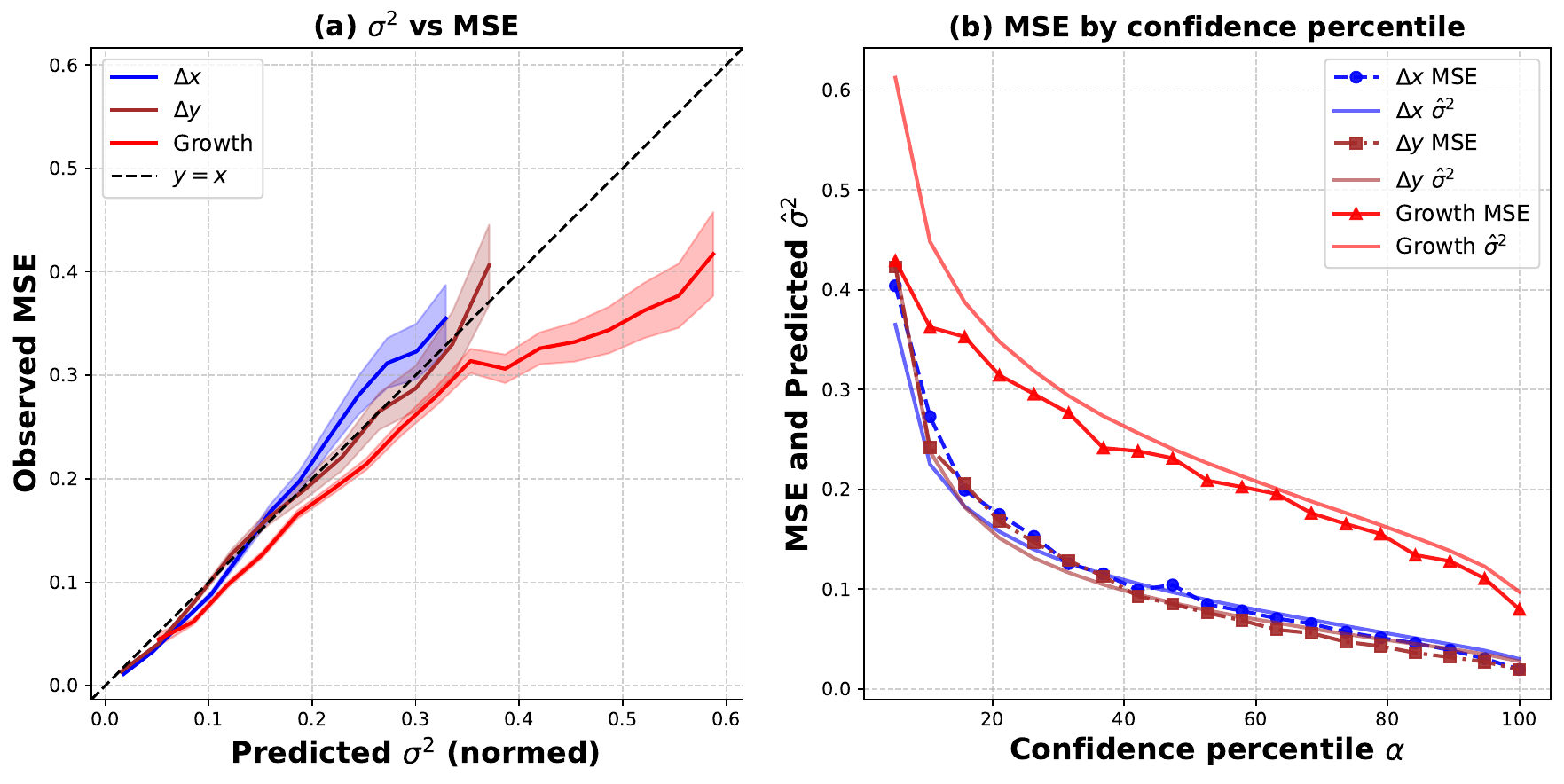}  
    \caption{(a) Mean MSE loss (calculated by variables normalized) of $\Delta x$, $\Delta y$, and growth as a function of the predicted (normalized) variance ($\boldsymbol{\sigma}^2$ of each predicted variable: $\Delta x$, $\Delta y$, and growth, respectively) at the 42th hour. Shading represents the ±1 standard error.
    (b) Mean observed MSE (black dashed curves with markers) and predicted variance $\hat{\boldsymbol{\sigma}}^2$ (solid curves) as a function of the confidence percentile $\alpha$ (higher $\alpha$ corresponds to higher network confidence, i.e., smaller predicted $\hat{\boldsymbol{\sigma}}^2$).}
    \label{fig:s3}
    
\end{figure}

\begin{figure}
    \centering
    \includegraphics[width=1.0\textwidth]{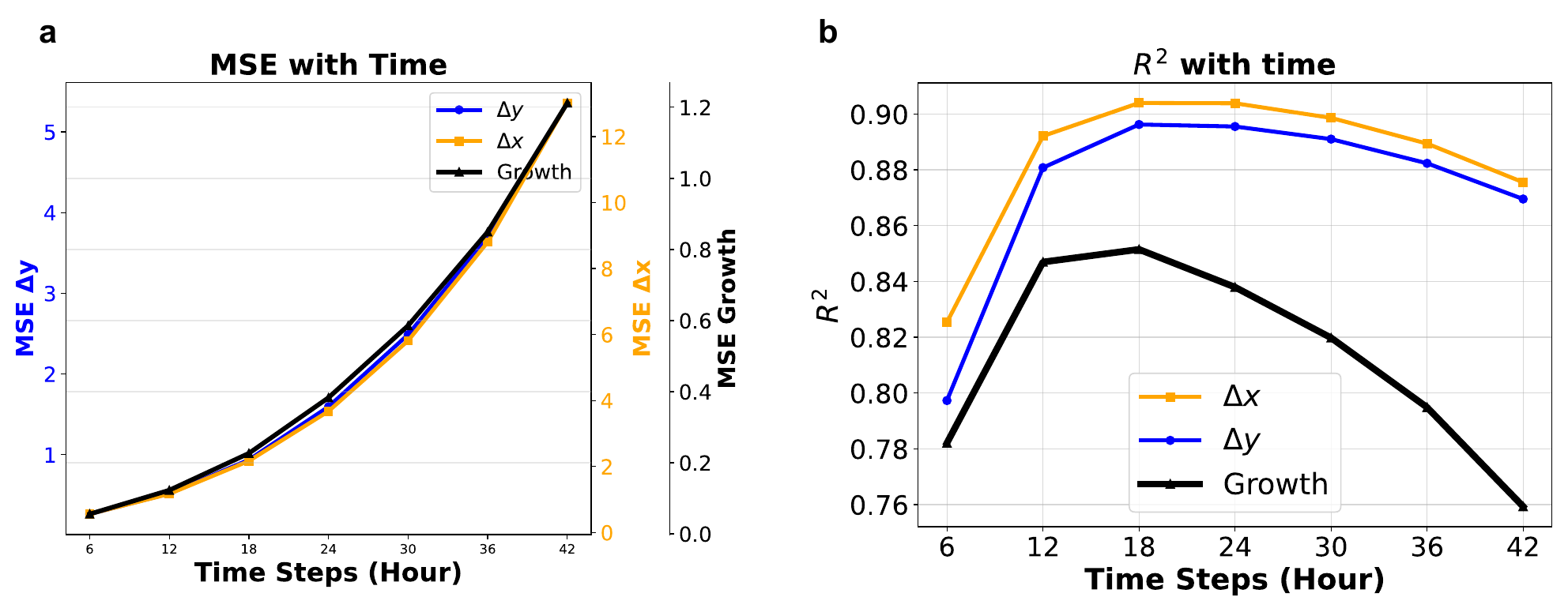}  
    \caption{(a) Mean squared error (MSE) of the CNN-predicted $\Delta x$ (unit: $^\circ$), $\Delta y$ (unit: $^\circ$) and "growth"  (unit: $10^{-5}\mathrm{s}^{-1}$); (a) Coefficient of determination $R^2$  of  the CNN-predicted $\Delta x$, $\Delta y$ and "growth", calculated at each prediction lead time step.  }
    \label{fig:s4}
    
\end{figure}

\begin{figure}
    \centering
    \includegraphics[width=0.9\textwidth]{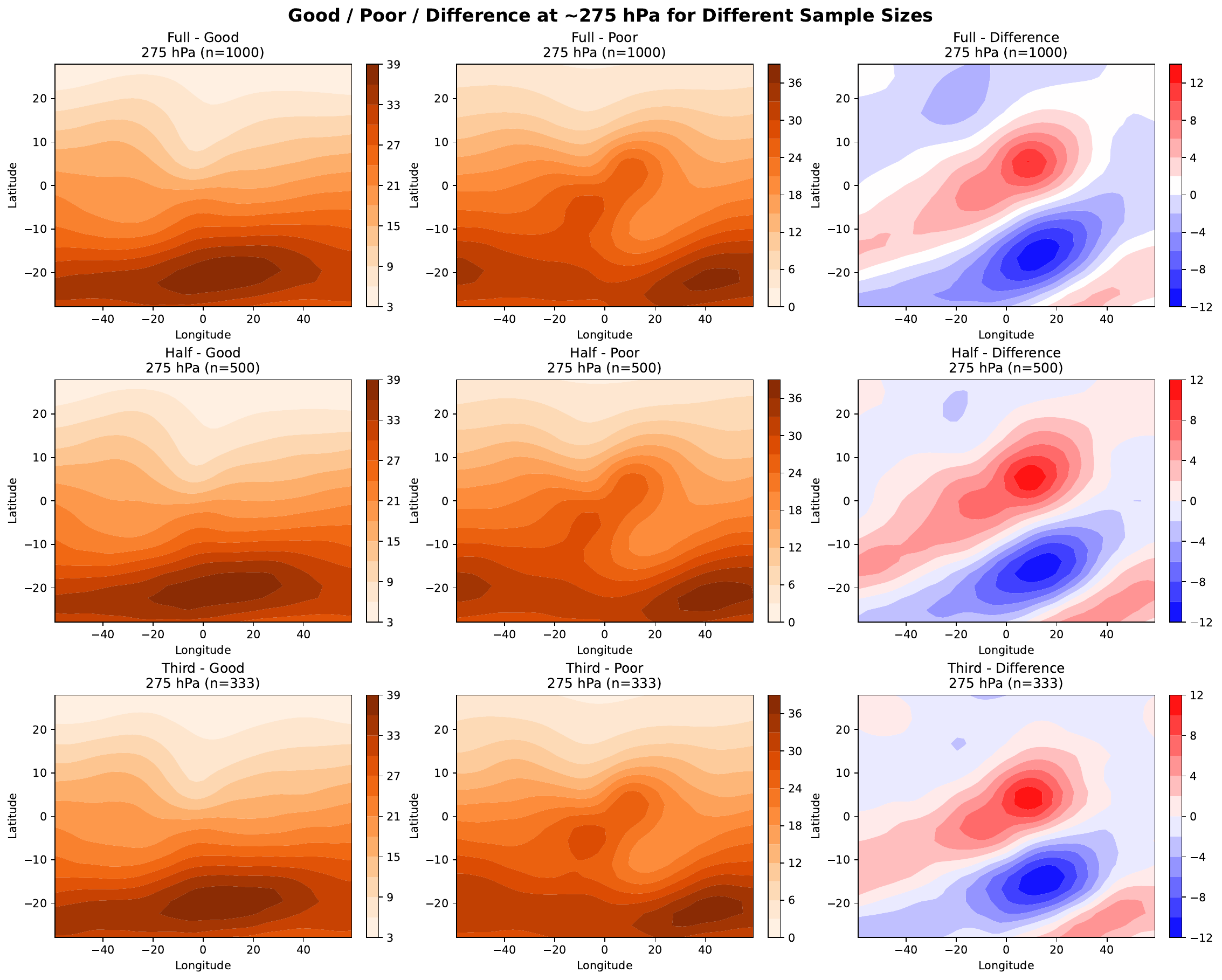}  
    \caption{Composite 275 hPa zonal wind fields at storm initialization for well-predicted storms (first column), poorly predicted storms (second column), and their difference (third column), for storms located north of 40$^\circ$ N. The first row is based on 1000 samples, the second row on 500 samples, and the third row on 333 samples. }
    \label{fig:s5}
    
\end{figure}

\begin{figure}
    \centering
    \includegraphics[width=1.0\textwidth]{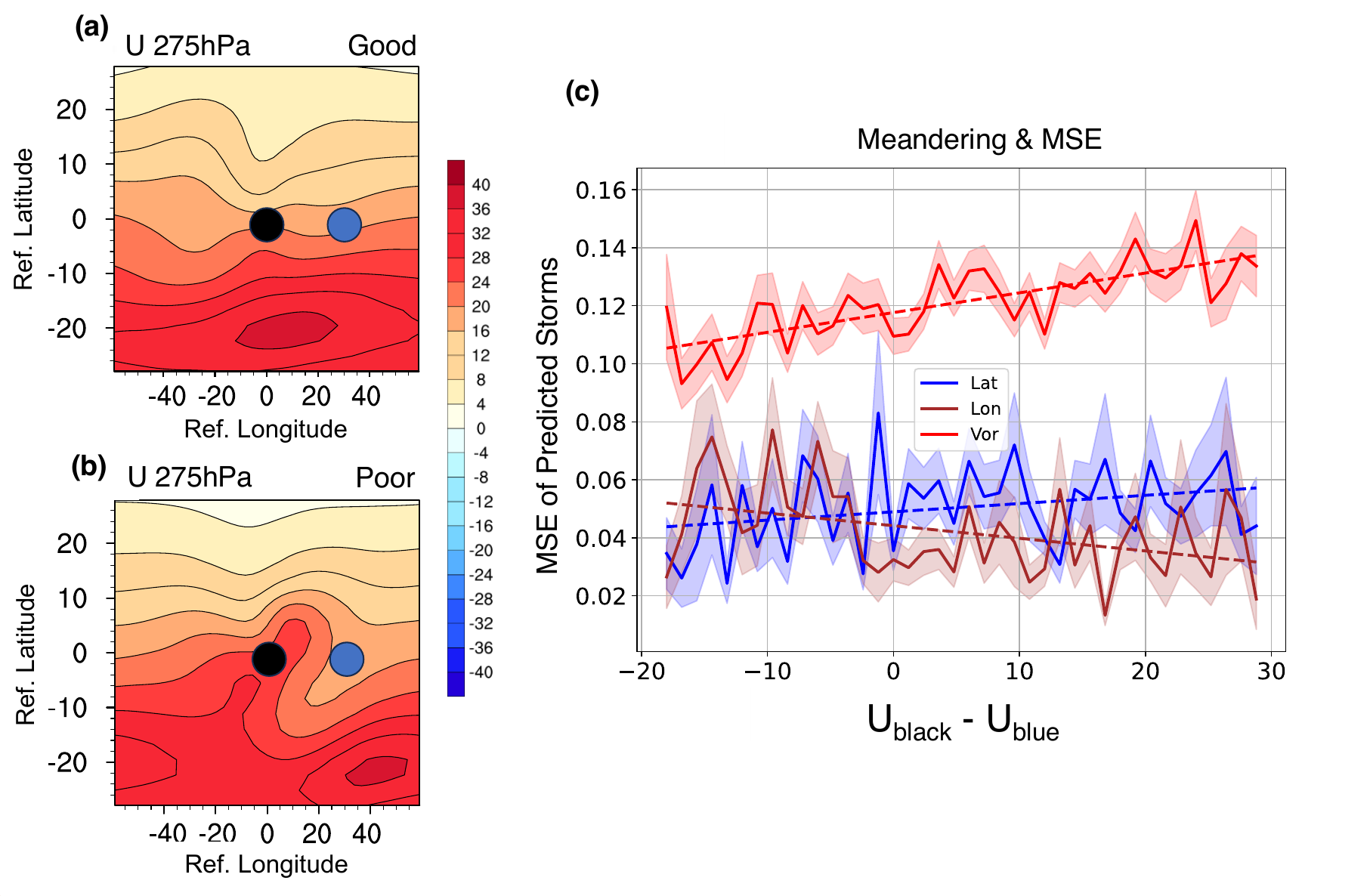}  
    \caption{(a-b) Composite initial 275 hPa zonal wind fields for (a) good and (b) poor predictions, for storms located north of 40$^\circ $N. (c) MSE of predicted storm intensity, latitude change and longitude change as a function of the Zonal-wind difference between 2 dots (as a quantification of du/dx in this "meandering" region of the poor) identified in (a-b).}
    \label{fig:s6}
    
\end{figure}

\begin{figure}
    \centering
    \includegraphics[width=1.0\textwidth]{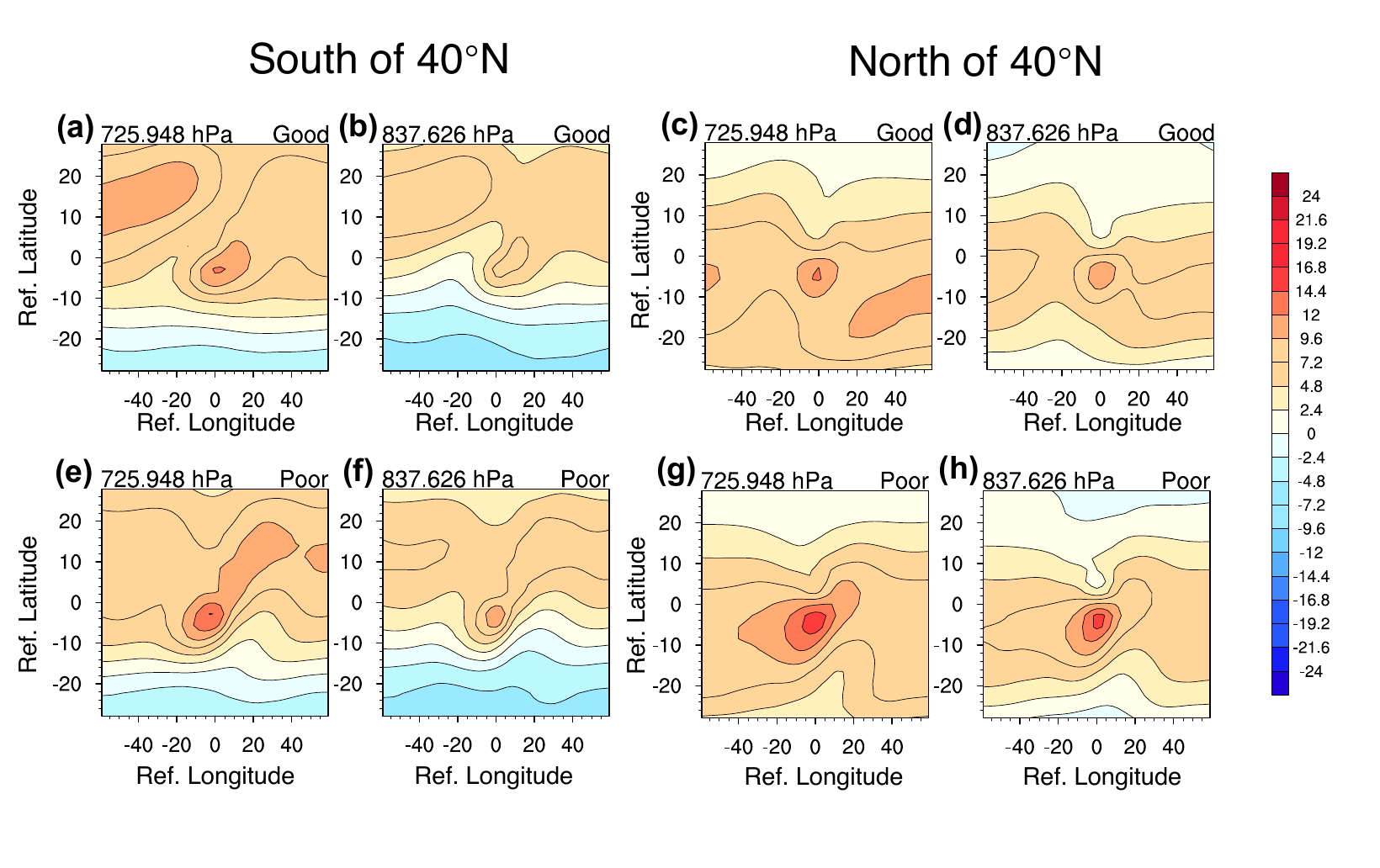}  
    \caption{Composite lower-level zonal wind fields at storm initialization for well- (a–d) and poorly-predicted (e–h) storms. Panels (a), (b), (e), and (f) show winds at 726 hPa; (c), (d), (g), and (h) show winds at 838 hPa. Panels (a), (b), (e), and (f) correspond to storms south of 40$^\circ$N, while (c), (d), (g), and (h) correspond to storms north of 40$^\circ$N. }
    \label{fig:s7}
    
\end{figure}

\begin{figure}
    \centering
    \includegraphics[width=0.6\textwidth]{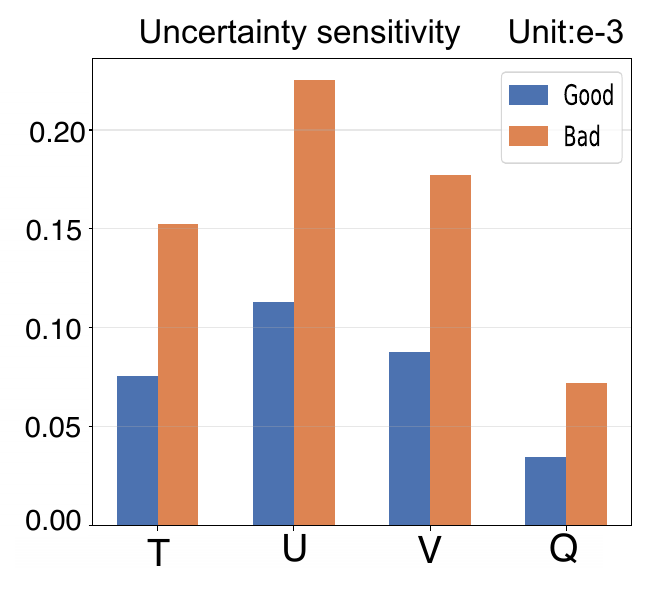}  
    \caption{Uncertainty sensitivity $\mathcal{E}(\mathbf{\chi})$, as defined in Equation 4, of the predicted 42h intensity change to input temperature, zonal-wind, meridional-wind and specific humidity, for both good and poor predictions (as classified in Figure 2b). We are averaging the sensitivity vectors by taking the vertical and horizontal mean of the absolute value of each vector component for comparison.}
    \label{fig:s8}
    
\end{figure}

\end{document}